\documentclass[11pt,twoside]{article}
\usepackage{amsfonts,amssymb,amsbsy}
\newcommand{\be}{\begin{equation}}
\newcommand{\ee}{\end{equation}}
\newcommand{\ba}{\hspace*{-5pt}\begin{array}}
\newcommand{\ea}{\end{array}}
\newcommand{\p}{\partial}
\newcommand{\bu}{\mathbf{u}}
\newcommand{\F}{\mathbf{F}}
\newcommand{\G}{\mathbf{G}}
\newcommand{\bH}{\mathbf{H}}

\def\Im{\mathop{\rm Im}\nolimits}

\newcommand{\om}{{\vec \omega}}
\newcommand{\ds}{\displaystyle}

\newtheorem{lem}{Lemma}

\newtheorem{prop}{Proposition}
\newtheorem{defi}{Definition}

\date{}
\author{Artur Sergyeyev\\
Silesian University in Opava,
Mathematical Institute,\\
Bezru\v{c}ovo n\'am.~13,
746~01 Opava,
Czech Republic\\
E-mail: Artur.Sergyeyev@math.slu.cz, arthurser@imath.kiev.ua}
\title{On recursion operators 
and nonlocal symmetries
of evolution equations\protect{\thanks{The research was
supported by Grants CEZ:J10/98:192400002 and VS 96003 ``Global Analysis'' 
of the Czech Ministry of Education, Youth and Sports and by the Grant 
No. 201/00/0724 of the Czech Grant~Agency.}}}
\begin{document}
\maketitle

\begin{abstract}
We consider the recursion operators with nonlocal terms of special form
for evolution systems in $(1+1)$ dimensions, and 
extend them
to well-defined operators 
on the space of nonlocal
symmetries associated with the so-called universal Abelian coverings over
these systems.
The extended recursion operators are shown to leave this space
invariant. These results apply, in particular, to the recursion operators
of the majority of known today
$(1+1)$-di\-men\-si\-on\-al integrable evolution systems.
We also present some related results and describe the extension
of them and of the above results to 
$(1+1)$-di\-men\-si\-on\-al systems
of PDEs transformable into the evolutionary form. 
Some examples and
applications are 
given.\looseness=-1 
\end{abstract}

\section{Introduction}
The scalar $(1+1)$-dimensional evolution
equation possessing an infinite-di\-men\-si\-on\-al commutative Lie
algebra of {\it time-independent} local generalized (Lie--B\"ack\-lund)
symmetries is usually either linearizable or integrable via inverse
scattering transform, see e.g.\
\cite{blaszak} -- \cite{ibrbook}, \cite{s,mik1,o,sok,wang}
for the survey of known results and \cite{mik} for the generalization to
(2+1) dimensions.
%
The existence of 
such algebra is usually proved by exhibiting the recursion operator \cite{o}
or master symmetry \cite{fu}. But in order to possess the latter the
equation in question must have 
higher order {\it time-dependent} symmetries, which usually turn out to be
nonlocal. This fact is one of the main reasons for growing interest in the
study of the whole algebra of time-dependent symmetries of evolution
equations~\cite{blaszak}. 
Moreover, for the evolution equations with
time-dependent coefficients
it is natural to consider their
time-dependent symmetries from the very beginning \cite{fu1, mbcf,svin}.
\looseness=-1

Nowadays there seem to exist 
two basic approaches to the definition
of nonlocal symmetries of PDEs.
They are presented in the papers of Fuchssteiner \cite{fu, fu1}
and the book of B\l aszak \cite{blaszak}, and references therein, 
and in the works of Vinogradov et al., see e.g.\ \cite{khor,v,v0} and references
therein.  Our approach is a combination of both.
\looseness=-1

In the present paper we shall consider the objects
called by Vinogradov et al.\ the {\em shadows} of
nonlocal symmetries associated with the so-called universal Abelian
covering (UAC) over the evolution system (\ref{eveq}). 
It was shown by Khor'kova \cite{khor} that in the case of UAC
any shadow can be lifted to a nonlocal symmetry over UAC in the sense of
\cite{khor, v}. Moreover, it is easy to see that the nonlocal
symmetries in the sense of \cite{khor, v} with zero shadows
are in a sense trivial and hardly represent any practical interest. Hence,
we lose no essential information about the nonlocal symmetries of
(\ref{eveq}), when we restrict ourselves to considering just 
the shadows.

Since most of authors, see e.g. \cite{blaszak,fu,fu1,svin} and
references therein, call similar objects just ``nonlocal symmetries", in
this paper we shall  essentially keep this tradition, calling the objects of
our study ``nonlocal UAC symmetries". This does not lead to any confusion,
because  nonlocal symmetries in the sense of \cite{khor, v} will not
appear in this paper. The precise definitions 
and further comments 
are given in Section~\ref{sec2} below.
Note that our interest in nonlocal UAC symmetries was inspired
by the results from \cite{khor, v}, where
it was shown that in certain cases it is possible to extend the action of 
the recursion operator so that its application to these symmetries produces
symmetries of the same kind.
\looseness=-1

In Section \ref{sec3} we prove two main results of the present paper.
The first one states that a large class of recursion operators for systems
(\ref{eveq}) can be extended to well-defined operators, namely to
recursion operators in the sense of Guthrie \cite{gut}, acting on the space
of nonlocal UAC symmetries and leaving this space invariant. The second
result states that under certain conditions, which are satisfied for the
majority of known examples, the repeated application of (extended)
recursion operator to weakly nonlocal UAC symmetries (i.e., to the
symmetries that
depend only on the nonlocal variables of the first level, that is, on the
integrals of local conserved densities, and are linear in these variables) 
again yields weakly nonlocal UAC symmetries. An important consequence of the
latter result is that the hereditary algebras (see e.g.\ \cite{blaszak}
for definition) of time-dependent symmetries for evolution
systems~(\ref{eveq}) are usually contained in  the set
of weakly nonlocal UAC symmetries. Note
that our results are  in a sense complementary to
Wang's~\cite{wang} sufficient conditions of {\em locality} of
time-independent symmetries, obtained with usage of recursion
operator.\looseness=-2 

In Section \ref{sec4} we 
explain how the results of Section~\ref{sec3} can be transferred from
$\hbox{(1+1)}$-dimensional evolution systems 
to arbitrary $(1+1)$-dimensional systems of PDEs transformable into the
evolutionary form by the appropriate change of variables. 
This is illustrated by the example of sine-Gordon equation. Finally, in
Section~\ref{sec5} we discuss further applications of our results and 
consider the example of Harry Dym equation.

In Appendices A and B we present two useful technical results: the
characterization of kernel of the total $x$-derivative in the space of
nonlocal UAC functions and the proof of the fact that the set of nonlocal UAC
symmetries for (\ref{eveq}) is a Lie algebra under the so-called Lie 
bracket---a natural commutator for nonlocal symmetries (cf.\ e.g.\
\cite{blaszak}). Note that if the hereditary algebra of time-dependent symmetries
for (\ref{eveq}) is generated by 
(a finite or infinite number of) 
nonlocal UAC symmetries,
then by virtue of the latter result  {\em all} elements of this algebra are
nonlocal UAC symmetries.

\section{Basic definitions and structures}\label{sec2}
Let us consider a $(1+1)$-dimensional system of evolution equations 
\begin{equation} \label{eveq}
 \partial \bu / \partial t =\F(x,t,\bu,\dots,\bu_{n})
 \end{equation}
for the $s$-component vector function $\mathbf{u}=(u^{1},\dots,u^{s})^{T}$.
Here
$\mathbf{u}_{j}=\partial^{j}\mathbf{u}/\partial
x^{j}$, $\mathbf{u}_{0}\equiv \mathbf{u}$; 
$\mathbf{F}=(F^{1},\dots, F^{s})^{T}$; $^{T}$ denotes the matrix transposition. 
As in \cite{muks,sok}, we make a {\em blanket assumption} that all
functions below ($\F$, symmetries $\G$, etc.) are locally analytic functions
of their arguments. This allows, in particular, to avoid the pathologies
caused by the existence of divisors of zero in the ring of
$C^{\infty}$-smooth functions.
\looseness=-2

\subsection{Universal Abelian covering} Following \cite{khor, v}, 
describe the construction of universal Abelian covering over a system of
PDEs for the particular case of system (\ref{eveq}).\looseness=-1 

A function $f(x,t,\bu,\bu_{1},\dots)$ is called {\em local}
(cf.\ \cite{s,mik1,mik}), if a) $f$ depends only on a finite number of
variables $\bu_k$ and b) $f$ is a locally analytic function of its
arguments.\looseness=-1  

The operators of total $x$- and $t$-derivatives on the space of local
functions are defined as (cf.\ e.g.\ \cite[Ch.\ V]{o})\looseness=-1
$$
\ba{l}
D_{x}^{(0)}=\p/\p x +
\sum\limits_{I=1}^{s}\sum\limits_{i=0}^{\infty}
 u^{I}_{i+1} \p/\p u_{i}^{I},\\
 D_{t}^{(0)}=\p/\p t +
 \sum\limits_{I=1}^{s}\sum\limits_{i=0}^{\infty}
 (D_{x}^{(0)})^{i}(F^{I})\p/\p u_{i}^{I}.
\ea
$$

Let
$\{D_{t}^{(0)}(\rho_{\alpha}^{(0)})=D_{x}^{(0)}(\sigma_{\alpha}^{(0)}) \mid
\alpha\in\mathcal{I}_1\}$ be a basis for the 
space $\mathrm{CL}_{F}^{(0)}$ of nontrivial local conservation laws for
(\ref{eveq}) considered modulo trivial ones. 
Recall that locality 
means that
$\rho_{\alpha}^{(0)}$ and $\sigma_{\alpha}^{(0)}$ are local functions 
and nontriviality means that
$\rho_{\alpha}^{(0)}\not\in\Im D_{x}^{(0)}$ (i.e.,
$\rho_{\alpha}^{(0)}$ cannot be represented as a total $x$-derivative
of a local function). 
\looseness=-1
%

We introduce \cite{khor, v} nonlocal variables $\omega_{\alpha}^{(1)}$ of
the first level as ``integrals" of $\rho_{\alpha}^{(0)}$. Namely, we define
them for all $\alpha\in\mathcal{I}_{1}$
as a solution of the system of PDEs\looseness=-1
\be\label{level1}
\ba{l}
 \partial \omega_{\alpha}^{(1)}/\partial x =\rho_{\alpha}^{(0)},\\[1mm]
\partial \omega_{\alpha}^{(1)}/\partial t =\sigma_{\alpha}^{(0)}.
\ea
\ee
%
It is clear that $\omega_{\alpha}^{(1)}$ are nothing but
the {\em potentials} for the conserved currents
$(\rho_{\alpha}^{(0)},\sigma_{\alpha}^{(0)})$.\looseness=-1

Now let us extend the action of operators $D_{x}^{(0)}$ and
$D_{t}^{(0)}$ to the functions that depend on
$\omega_{\alpha}^{(1)}$ by means
of the formulae 
$$
\ba{l}
D_{x}^{(1)}=D_{x}^{(0)}+\sum\limits_{\alpha\in\mathcal{I}_{1}}
\rho_{\alpha}^{(0)}\p/\p \omega_{\alpha}^{(1)},\quad 
D_{t}^{(1)}=D_{t}^{(0)}+\sum\limits_{\alpha\in\mathcal{I}_{1}}
\sigma_{\alpha}^{(0)}\p/\p \omega_{\alpha}^{(1)}
\ea
$$
and consider a basis $\{D_{t}^{(1)}(\rho_{\alpha}^{(1)})=
D_{x}^{(1)}(\sigma_{\alpha}^{(1)})\mid \alpha\in\mathcal{I}_2\}$ in the set
$\mathrm{CL}_{F}^{(1)}$ of all 
nontrivial (nontriviality now means that
$\rho_{\alpha}^{(1)}\not\in\Im D_{x}^{(1)}$) local conservation laws for
the system of equations (\ref{eveq}), (\ref{level1}).
The densities $\rho_{\alpha}^{(1)}$ and fluxes
$\sigma_{\alpha}^{(1)}$ may 
depend not only on
$x,t,\bu$, $\bu_{1},\bu_{2},\dots$, but
on $\om^{(1)}$ as well, and any given $\rho_{\alpha}^{(1)}$ or
$\sigma_{\alpha}^{(1)}$ depends only on a finite number of
$\bu_{r}$ and of nonlocal variables $\omega_{\alpha}^{(1)}$.
Here $\om^{(1)}$ denotes the totality of
variables $\omega_{\alpha}^{(1)}$ for $\alpha\in\mathcal{I}_{1}$. 
\looseness=-2

We further define the nonlocal variables of second level by means of
the relations
$$
\ba{l}
 \partial \omega_{\alpha}^{(2)}/\partial x =
 \rho_{\alpha}^{(1)},\alpha\in\mathcal{I}_{2},\\[3pt]
\partial \omega_{\alpha}^{(2)}/\partial t
=\sigma_{\alpha}^{(1)},\alpha\in\mathcal{I}_{2},
\ea
$$
extend the action of $D_{x}^{(1)}$ and $D_{t}^{(1)}$
to the functions that may depend on $\omega_{\alpha}^{(2)}$, and so
on.\looseness=-1

Iterating this procedure infinite number of times, we obtain an
infinite-dimensional covering $\mathcal{U}$ over (\ref{eveq}), which is
called {\em universal Abelian covering} (UAC) \cite{khor,v}. 
More precisely, the covering constructed in this way is
just 
a representative of the class of equivalent coverings,
and the authors of \cite{khor, v} identify UAC with this
class.\looseness=-1

Thus, 
$\mathcal{U}$
involves the infinite set of nonlocal
variables
$\omega_{\alpha}^{(j)}$ defined by 
the relations
\begin{eqnarray}
\p\omega_{\alpha}^{(j)}/\p x=\rho_{\alpha}^{(j-1)},
\alpha\in\mathcal{I}_{j},
j\in\mathbb{N},\label{cov1}\\
\p\omega_{\alpha}^{(j)}/\p t=\sigma_{\alpha}^{(j-1)},
\alpha\in\mathcal{I}_{j},j\in\mathbb{N}.\label{cov2}
\end{eqnarray}
Here $\mathcal{I}_{k+1}$,
$k\geq 1$, is a set of indices such that the conservation laws
$D_{t}^{(k)}(\rho_{\alpha}^{(k)})\allowbreak=D_{x}^{(k)} (\sigma_{\alpha}^{(k)})$
for
$\alpha\in\mathcal{I}_{k+1}$ form a basis in the set
$\mathrm{CL}_{F}^{(k)}$ of all  nontrivial local conservation laws 
of the form $D_{t}^{(k)}(\rho)=D_{x}^{(k)}(\sigma)$ for (\ref{eveq}) and
(\ref{cov1}), (\ref{cov2}) with $j\leq k$. The locality means that the   
densities
$\rho$ and fluxes $\sigma$ of conservation laws from
$\mathrm{CL}_{F}^{(k)}$ depend only on
$x,t,\bu,\bu_{1},\dots$ and $\om^{(1)},
\dots,\om^{(k)}$, but
not on $\om^{(m)}$ with $m>k$, and any given density or flux
depends only on a finite number of $\bu_{r}$ and of nonlocal variables
$\omega_{\alpha}^{(j)}$. The nontriviality of a conservation law 
$D_{t}^{(k)}(\rho)=D_{x}^{(k)}(\sigma)$ from
$\mathrm{CL}_{F}^{(k)}$ means that $\rho$ cannot be represented in the form
$D_{x}^{(k)}(f)$ for some
$f=f(x,t,\om^{(1)},\dots,\om^{(k)},\bu,\bu_1,\dots)$.
%
\looseness=-1

We employed here the notation
$$
\ba{l}
D_{x}^{(k)}=D_{x}^{(k-1)}+\sum\limits_{\alpha\in\mathcal{I}_{k}}
\rho_{\alpha}^{(k-1)}\p/\p \omega_{\alpha}^{(k)},\quad 
D_{t}^{(k)}=D_{t}^{(k-1)}+\sum\limits_{\alpha\in\mathcal{I}_{k}}
\sigma_{\alpha}^{(k-1)}\p/\p \omega_{\alpha}^{(k)},
\ea
$$
and the notation $\om^{(k)}$ for the totality of variables
$\omega_{\alpha}^{(k)}$,
$\alpha\in\mathcal{I}_{k}$. We shall also denote by $\om$ 
the totality of variables $\omega_{\alpha}^{(j)}$ for all $j$ and $\alpha$.

The operators of total derivatives on the space of functions of
$x,t,\om,\bu,\bu_1,\dots$
are
$$ \ba{l} D\equiv D_{x}=D_{x}^{(0)}+
\sum\limits_{j=1}^{\infty}\sum\limits_{\alpha\in\mathcal{I}_{j}}
\rho_{\alpha}^{(j-1)}\p/\p \omega_{\alpha}^{(j)},\\
D_{t}=D_{t}^{(0)}+
\sum\limits_{j=1}^{\infty}\sum\limits_{\alpha\in\mathcal{I}_{j}}
\sigma_{\alpha}^{(j-1)}\p/\p \omega_{\alpha}^{(j)}. \ea $$

The relations
$D_{t}^{(k)}(\rho_{\alpha}^{(k)})=D_{x}^{(k)}(\sigma_{\alpha}^{(k)})$
imply the compatibility of (\ref{cov1}) and (\ref{cov2}). In turn, the
consequence of the latter and of the equality $\lbrack
D_{x}^{(0)},D_{t}^{(0)}\rbrack=0$ are the relations
$$
\ba{l}
\lbrack D_{x}^{(k)},D_{t}^{(k)}\rbrack=0, k=1,2,\dots,\\
\lbrack D_{x},D_{t}\rbrack=0.
\ea
$$

We shall say (cf.\ \cite{s, mik, muks}) that a function
$f=f(x,t,\om,\allowbreak\bu,\bu_1,\dots)$ 
is a {\em nonlocal UAC} function, if a) $f$
depends only on a finite number of variables
$\omega_{\alpha}^{(j)}$ and $\bu_{k}$ and b) $f$ is a locally analytic
function of its arguments.  
We shall call $f$ a nonlocal UAC function of level
$k$, if $f$ is a nonlocal UAC function 
independent of 
$\omega_{\alpha}^{(j)}$ for $j>k$.\looseness=-2

Since the kernel of $D$ 
in the space of nonlocal UAC functions is
exhausted by functions of~$t$ (see Appendix A for the proof), it is easy to
verify that our definition  (cf.\ \cite{mik1}) of
nontriviality of a conservation law is in fact equivalent to the standard
one \cite{o, v}. 

Let us stress that $\rho_{\alpha}^{(k-1)}$ are defined up to
the addition of the terms from $\Im D_{x}^{(k-1)}$, and
$\sigma_{\alpha}^{(k-1)}$ are defined up to
the addition of the terms from $\ker D_{x}^{(k-1)}$,
i.e., they
should be considered as equivalence classes modulo
$\Im D_{x}^{(k-1)}$ and $\ker D_{x}^{(k-1)}$, respectively.
This means, in particular, that the nonlocal variables
$\omega_{\alpha}^{(k)}$ are defined up to the addition of
arbitrary nonlocal UAC functions of level $k-1$.

Making different choices of representatives in these equivalence
classes yields different coverings over (\ref{eveq}), but these
coverings are equivalent in the sense of definition from \cite[Ch.\ 6]{v},
and in the sequel
we shall assume that we deal with a fixed
representative of the respective equivalence class, because
constantly operating with the whole class in the explicit computations
is extremely inconvenient. However,
the results obtained below are obviously independent of this
choice, and thus hold true for the whole class of equivalent
coverings, which result from the above construction.\looseness=-1


%
%
\subsection{Nonlocal UAC symmetries}
We shall call (cf.\ \cite{blaszak,muks,o,v}) an $s$-component nonlocal
UAC vector function $\G$ a {\em nonlocal UAC symmetry} of (\ref{eveq}), if
the evolution system $\p\bu/\p{\tau}=\G$ is compatible with (\ref{eveq}),
i.e., $\p^{2}\bu/\p t\p\tau=\p^{2}\bu/\p\tau\p t$, where the derivatives
with respect to $t$ and $\tau$ are computed with usage of
(\ref{eveq}), (\ref{cov2}) and $\p\bu/\p{\tau}=\G$, respectively.
We shall denote the set of all nonlocal UAC symmetries
of (\ref{eveq}) by $\mathrm{NS}_{F}(\mathcal{U})$. 
If $\p\G/\p\om=0$, then
$\G$ is called (local) {\em generalized} (or {\em higher local}, or just 
{\em local}) sym\-metry of (\ref{eveq}), see e.g.\ \cite{s,mik,o}.
\looseness=-1
With $\G$ being a nonlocal UAC vector function, the compatibility condition
for (\ref{eveq}) and $\p\bu/\p{\tau}=\G$ takes the form
\be\label{symql}
D_{t}(\G)=\F'[\G],
\ee
where $\F'=\sum
_{i=0}^{n}\p\F/\p\bu_{i}D^{i}$.

Let us mention that nonlocal UAC symmetries
are nothing but a particular case of general nonlocal
symmetries, considered e.g.\ in \cite{blaszak,fu}. Indeed, the determining
equations for the latter, given in \cite{blaszak,fu}, are
nothing but the compatibility conditions for (\ref{eveq}) and
$\bu_{\tau}=\G$, and these conditions reduce to (\ref{symql}),
if $\G$ is a nonlocal UAC vector function.\looseness=-1

The set $\mathrm{NS}_{F}(\mathcal{U})$ of nonlocal UAC symmetries is
interesting and important. In particular,
our results imply that 
for nearly all known examples 
the action
of the recursion operator is well defined on $\mathrm{NS}_{F}(\mathcal{U})$ 
and leaves it
invariant. Hence, $\mathrm{NS}_{F}(\mathcal{U})$ 
contains all elements of the hereditary algebra
(see \cite{blaszak,fu,fu1} for its precise definition) of time-dependent
symmetries for (\ref{eveq}),
if they are generated (cf.\ e.g.\ \cite{blaszak}) by means of the
repeated application of the recursion operator  
to the scaling symmetry
of (\ref{eveq}) and to a time-independent local generalized symmetry of
(\ref{eveq}) (e.g.\ to $\F$, if $\p\F/\p t=0$).  
Note that $\mathrm{NS}_{F}(\mathcal{U})$ is a Lie algebra with respect to the
so-called Lie bracket (see 
Appendix B below for the proof). Therefore, if the hereditary
algebra for (\ref{eveq}) is generated by (a finite or infinite 
number of) nonlocal UAC symmetries, then {\em all} its elements 
are nonlocal UAC symmetries. 
Note that there also exist (see e.g.\ \cite{s,
svin}) integrable systems (\ref{eveq})
that possess only a finite number of
local generalized symmetries, but have infinite hierarchies of nonlocal
symmetries, and 
these nonlocal symmetries
turn out to be nonlocal UAC ones.\looseness=-2

To avoid possible confusion, let us stress that the definition
of nonlocal symmetries used in \cite{khor, v} is
different from the ours, but 
by Theorem~3.1 from \cite{khor} 
we always can recover from 
$\G\in \mathrm{NS}_{F}(\mathcal{U})$
the nonlocal
symmetry in the sense of
\cite{khor, v}.\looseness=-1

\section{On the action of recursion operators}\label{sec3}
In this section we prove for
the 
case of systems (\ref{eveq}) and recursion
operators of the form (\ref{recop}) the conjecture of Khor'kova
\cite{khor} stating that for any $(1+1)$-dimensional system of PDEs 
its recursion operator $\mathfrak{R}$ can be extended to a well-defined
operator $\tilde{\mathfrak{R}}$
on the space $\mathrm{NS}_{F}(\mathcal{U})$ 
of nonlocal UAC symmetries 
and leaves this space invariant.

In complete analogy with the 
case of local functions, see e.g.\
\cite[Ch.\ V, \S 5.3]{o}, we call the operator 
$\mathfrak{B}^{\dagger}=\sum\limits_{i=0}^{q}(-D)^{i}\circ
b_{i}^{T}$
a  {\em formal adjoint} 
of 
$\mathfrak{B}=\sum\limits_{i=0}^{q}b_{i}D^{i}$. Here $b_{i}$ are
some $r\times r$ matrix-valued nonlocal UAC functions, 
$^{T}$ denotes the matrix transposition and $\circ$
stands for the composition of operators.\looseness=-1

An $s$-component nonlocal UAC vector function
$\boldsymbol{\gamma}$ is called a {\em cosymmetry} \cite{blaszak,wang}
of~(\ref{eveq}), if it satisfies the equation
\be\label{cosym}
D_{t}(\boldsymbol{\gamma})+(\F')^{\dagger}[\boldsymbol{\gamma}]=0.
\ee 

Nearly all known today recursion operators for systems (\ref{eveq}) have the
form (cf.\ 
\cite{wang})\looseness=-1
\be\label{recop}
\mathfrak{R}=\sum\limits_{i=0}^{k}
a_{i}D^{i}+\sum\limits_{j=1}^{p} \G_{j} \otimes D^{-1}\circ
\boldsymbol{\gamma}_{j}, \ee where $a_{i}$ are $s\times s$
matrix-valued local functions, and $\G_{j}$ and
$\boldsymbol{\gamma}_{j}$ 
are 
local symmetries
and cosymmetries of (\ref{eveq}), respectively. 

Note that the action of recursion operators of the form (\ref{recop}) is
initially defined only on {\em local} generalized symmetries (see e.g.\
\cite{o}), but 
the formula (\ref{recop}) together 
with the subsequent definition of $D^{-1}$ enable us to 
construct an extension $\tilde{\mathfrak{R}}$  
of $\mathfrak{R}$ to the space $\mathrm{NS}_{F}(\mathcal{U})$. 
However, it is not clear {\em a priori} whether 
$\tilde{\mathfrak{R}}$ is a well-defined operator on
$\mathrm{NS}_{F}(\mathcal{U})$  and whether $\tilde{\mathfrak{R}}(\G)$ 
is a nonlocal UAC symmetry of (\ref{eveq}), provided so is
$\G$.\looseness=-1  

Let us mention the following result of Wang \cite{wang}. Suppose that
$\mathfrak{R}$ (\ref{recop}), with $\G_j$ and $\boldsymbol\gamma_{j}$
being arbitrary
$s$-component time-in\-de\-pen\-dent local vector functions, is a 
recursion operator for (\ref{eveq}), both $\mathfrak{R}$ and system
(\ref{eveq}) 
are homogeneous with respect
to a scaling of 
$x,t$ and $\bu$, and $\p\F/\p t=0$ and $\p\mathfrak{R}/\p t=0$. 
Then 
under 
minor 
restrictions on $\F$ and
$\mathfrak{R}$ the functions $\G_{j}$ and $\boldsymbol\gamma_{j}$ indeed
are symmetries and cosymmetries for (\ref{eveq}), respectively.
Obviously, 
this result remains valid when $a_{i}$,
$\G_{j}$ and
$\boldsymbol{\gamma}_{j}$ are nonlocal UAC functions.\looseness=-2

In order to proceed, we should agree how to interpret the action of $D^{-1}$.
For our purposes it suffices to adopt the
following definition, which is a particular case of the general construction
of Guthrie \cite{gut}:
\begin{defi}\label{defd-1}
Let $P$ be a nonlocal UAC function such that $D_{t}(P)=D(Q)$ for 
some nonlocal UAC function $Q$ (i.e., $P$ is a conserved density).

Then we shall understand under $R=D^{-1}(P)$ a solution of the system
\be\label{invd}
\ba{l}
D(R)=P,\\
D_{t}(R)=Q.
\ea
\ee
\end{defi}

The existence of nonlocal UAC solution $R$ of (\ref{invd}),
provided $D_{t}(P)=D(Q)$, is one of the fundamental
properties of universal Abelian covering, proved in \cite{khor}. 
Note that according to the above definition we have 
${\omega}_{\alpha}^{(j)}=D^{-1} (\rho_{\alpha}^{(j-1)})$, 
as it would be natural to expect. As we show in Appendix A,
the kernel of $D$ in the space of nonlocal UAC functions is
exhausted by functions of $t$, whence it is immediate that the solution $R$
of~(\ref{invd}) for given $P$ and $Q$ is unique up to the addition of an
arbitrary constant. So, the ``integral'' $D^{-1}(P)$, defined above,
is not an integral of $P$ itself, but rather of a conservation
law $D_{t}(P)=D(Q)$. 
It is also clear that the integrals $D^{-1}(P)$ evaluated for different
$Q$ in general will differ by a function of $t$. 
\looseness=-1 

In order to prove that we can extend the action of $\mathfrak{R}$ to any
nonlocal UAC symmetry $\bH$ of (\ref{eveq}), 
we have to show that 
for any $\bH\in\mathrm{NS}_{F}(\mathcal{U})$ we have 
$D_{t}(\boldsymbol{\gamma}_{j}\bH)=D(\zeta_j)$ for some nonlocal UAC
functions $\zeta_j$. Indeed, then 
the integrals
$D^{-1}({\boldsymbol{\gamma}}_{j}\bH)$, interpreted in the sense of the
above definition with $Q=\zeta_j$, are nonlocal UAC functions 
defined up to the addition of arbitrary constants. 
Note that the expressions like
$\mathbf{a}\mathbf{b}$ stand here and below for the scalar product of
$s$-component  vectors $\mathbf{a}$~and~$\mathbf{b}$.\looseness=-1
\looseness=-1 

By (\ref{symql}) and (\ref{cosym}) 
we have $D_{t}(\bH)=\F'[\bH]$ and
$D_{t}(\boldsymbol{\gamma}_{j})=
-(\F')^{\dagger}[\boldsymbol{\gamma}_{j}]$, whence 
\be\label{dtrhs}
D_{t}(\boldsymbol{\gamma}_{j}\bH)=
-(\F')^{\dagger}[\boldsymbol{\gamma}_{j}]\bH
+\boldsymbol{\gamma}_{j}\F'[\bH]. 
\ee

There is an obvious generalization (cf.\ e.g.\ \cite[Ch.\ V, \S
5.3]{o}) of the well-known Lagrange
identity from theory of ordinary differential equations, namely
\be\label{lagr} 
\ba{l}
{\vec{f}}\mathfrak{B}({\vec{g}})-
\mathfrak{B}^{\dagger}({\vec{f}}){\vec{g}}= D(\eta),\quad 
\eta=\sum\limits_{i=1}^{q}\sum\limits_{j=0}^{i-1}(-D)^{j}(b_{i}^{T}
{\vec{f}})D^{i-j-1}({\vec{g}}),
\ea
\ee
valid for any differential operator
$\mathfrak{B}= \sum\limits_{i=0}^{q}b_{i}D^{i}$ and for any 
$r$-com\-ponent nonlocal UAC vector functions ${\vec{f}}$ and ${\vec{g}}$,
provided $b_{i}$ are $r \times r$ matrix-valued nonlocal UAC
functions. \looseness=-1 

Using (\ref{lagr}) for $\mathfrak{B}=\F'$, $\vec
f=\boldsymbol{\gamma}_{j}$, $\vec g=\bH$, we conclude that
$D_{t}(\boldsymbol{\gamma}_{j}\bH)$
indeed
can be represented in the form
$D(\zeta_{j})$ for the nonlocal UAC function 
\be\label{lagr3}
\zeta_{j}=\sum\limits_{i=1}^{n}\sum\limits_{m=0}^{i-1}(-D)^{m}(
(\p\F/\p\bu_{i})^{T}\boldsymbol{\gamma}_{j})D^{i-m-1}(\bH).
\ee

Hence, for $P=\boldsymbol{\gamma}_{j}\bH$ we can always make a
`canonical' choice $Q=\zeta_{j}$ while computing $D^{-1}(P)$ according to
Definition~\ref{defd-1}.
With this choice and the above definition of $D^{-1}$, 
the recursion operator $\mathfrak{R}$ (\ref{recop})
is, in essence, replaced by a new operator 
$\tilde\mathfrak{R}$, which is easily seen to be a recursion operator
in the sense of Guthrie \cite{gut}. 

Many important properties and definitions (for
instance, that of hereditarity) can be readily transferred from
$\mathfrak{R}$ to
$\tilde{\mathfrak{R}}$. 
%
The operator $\tilde\mathfrak{R}$ is free \cite{gut} of the pathologies 
caused by na\"\i{}ve definition of $D^{-1}$, cf.\ \cite{sw} for
an alternative way of overcoming these difficulties. In particular, 
it is easy to see that $\tilde\mathfrak{R}$ always maps nonlocal UAC
symmetries to nonlocal UAC symmetries (cf.~\cite{gut}), because
$\mathfrak{R}$ 
satisfies the 
equation $[D_{t}-\F',\mathfrak{R}]=0$ \cite{o}. 

Thus,
we have proved\looseness=-1
\begin{prop}\label{rect1}
Any recursion operator $\mathfrak{R}$ 
(\ref{recop}) for (\ref{eveq}), with 
$a_{i}$ being $s\times s$
matrix-valued nonlocal UAC functions, 
and $\G_{j}$ and $\boldsymbol{\gamma}_{j}$
being nonlocal UAC symmetries and cosymmetries for (\ref{eveq}),
respectively, 
can be extended to a well-defined operator $\tilde{\mathfrak{R}}$ 
that acts
on the whole space 
$\mathrm{NS}_{F}(\mathcal{U})$
of nonlocal UAC symmetries
for (\ref{eveq}) and leaves this space
invariant.\looseness=-1
\end{prop}

Now let us consider what happens if $a_{i},\G_{j},
\boldsymbol{\gamma}_{j}$ are local and we apply $\tilde\mathfrak{R}$ to
a local generalized symmetry $\bH$ of (\ref{eveq}). The above
reasoning indeed holds true. Moreover, we see that
$\boldsymbol{\gamma}_{j}\bH$ are {\em local} conserved densities for
(\ref{eveq}) and the respective $\zeta_{j}$ are local as well. Hence, the
application of $\tilde{\mathfrak{R}}$ to local generalized symmetries of
(\ref{eveq}) yields nonlocal UAC symmetries 
of the form 
(cf.\ \cite{khor} and \cite[Ch.\ 6]{v})\looseness=-1
\be\label{weaknl}
{\bH}=\bH_{0}+\sum\limits_{\alpha\in\mathcal{I}_{H}}
\bH_{\alpha}\omega_{\alpha}^{(1)}, \ee 
where $\bH_{0}$ and $\bH_{\alpha}$ are $s$-component local vector functions,
and $\mathcal{I}_{H}$ is a finite subset of~$\mathcal{I}_{1}$. 

We shall call an $r$-component nonlocal UAC vector function {\em weakly
nonlocal UAC} vector function, if it can be represented in the form 
(\ref{weaknl}) with $\bH_{0}$ and $\bH_{\alpha}$ being $r$-com\-po\-nent
local vector functions. We shall denote by $\mathrm{WNLS}_{F}(\mathcal{U})$
the set of all weakly nonlocal UAC symmetries 
for (\ref{eveq}).
For the majority of integrable systems~(\ref{eveq})
their master symmetries are $s$-component weakly nonlocal UAC vector
functions.
%
\looseness=-1

Note that if 
$\bH$ is a local generalized symmetry of
(\ref{eveq}) and 
\be\label{locs}
\boldsymbol{\gamma}_{j}\bH=D(\xi_{j}),
\ee
where $\xi_{j}$ are local functions, then by
the above $D^{-1}(\boldsymbol{\gamma}_{j}\bH)$ can differ from $\xi_{j}$
only by a function of $t$. Hence,
$D^{-1}(\boldsymbol{\gamma}_{j}\bH)$ is a local function, and thus
$\tilde\mathfrak{R}(\bH)$ is a {\em local} generalized symmetry for
(\ref{eveq}). In other words, the application of $\tilde{\mathfrak{R}}$ to
local generalized symmetries of (\ref{eveq}) satisfying (\ref{locs})
again yields local generalized symmetries of (\ref{eveq}). 
Below we shall assume (obviously without loss of generality) that $\G_j$ in
(\ref{recop}) are linearly independent. Then it is easy to see that the
conditions (\ref{locs}) for $j=1,\dots,p$ are {\em equivalent} to the
requirement that  $\tilde{\mathfrak{R}}(\bH)$ is a local generalized
symmetry, provided so is $\bH$.
Let us mention that Theorems 6--8 and 6--9 of Wang \cite{wang} provide an
easy way to verify the conditions (\ref{locs}) for large
families of time-independent local generalized symmetries of (\ref{eveq}).
\looseness=-1

Since $\F$ and the coefficients of $D_t^{(1)}$ are independent of
$\omega_{\alpha}^{(j)}$ for all $\alpha$ and $j$, it is immediate that for
any nonlocal UAC symmetry $\G$ of level one for (\ref{eveq}) the quantities
$\p\G/\p\omega_{\alpha}^{(1)}$ satisfy the determining equation
(\ref{symql}) and hence also are nonlocal UAC symmetries of level
one for (\ref{eveq}). In particular, for any $\bH$ of the form
(\ref{weaknl}) the quantities
$\bH_{\alpha}=\p\bH/\p\omega_{\alpha}^{(1)}$ are in fact local generalized
symmetries of (\ref{eveq}).\looseness=-1

Using this result, let us show that 
$\tilde{\bH}=\tilde{\mathfrak{R}}(\bH)\in\mathrm{WNLS}_{F}(\mathcal{U})$ for
any
$\bH$ of the form (\ref{weaknl}), provided  
$\tilde{\mathfrak{R}}(\bH_{\alpha})$ are {\em local} generalized symmetries
of (\ref{eveq}) (or, equivalently, 
$\boldsymbol{\gamma}_{j}\bH_{\alpha}= D(\xi_{j,\alpha})$ for all
$j=1,\dots,p$ and all $\alpha\in\mathcal{I}_{H}$, where
$\xi_{j,\alpha}$ are local functions).\looseness=-1

As $D^{i}(\omega_{\alpha}^{(1)})=D^{i-1}(\rho_{\alpha}^{(0)})$ are local
functions for
$i\geq 1$, it is clear that  
$\tilde{\mathfrak{R}}(\bH)\in\mathrm{WNLS}_{F}(\mathcal{U})$,
if there exist scalar weakly nonlocal  UAC
functions $R_j$ such that
\begin{eqnarray}
D(R_{j})=\boldsymbol{\gamma}_{j}\bH,\label{rx}\\
D_{t}(R_{j})=\zeta_{j},\label{rt}
\end{eqnarray}
where $\zeta_{j}$ are given by (\ref{lagr3}). Indeed, then
$\tilde{\mathfrak{R}}(\bH)$ is a weakly
nonlocal UAC vector function, and by Proposition~\ref{rect1}
$\tilde{\mathfrak{R}}(\bH)\in\mathrm{NS}_{F}(\mathcal{U})$,
hence $\tilde{\mathfrak{R}}(\bH)\in\mathrm{WNLS}_{F}(\mathcal{U})$.

Let 
$\tilde{R}_{j}=R_{j}-\smash{\sum\limits_{\alpha\in\mathcal{I}_{H}}}
\xi_{j,\alpha}\omega_{\alpha}^{(1)}$. Using (\ref{rx}), (\ref{rt}), we
obtain
$$
\ba{l}
D(\tilde{R}_{j})=\boldsymbol{\gamma}_{j}\bH_{0}-
\sum\limits_{\alpha\in\mathcal{I}_{H}}\xi_{j,\alpha}\rho_{\alpha}^{(0)}
\equiv\psi_j,
\label{rx1}\\
D_{t}(\tilde{R}_{j})=\zeta_{j}-
\sum\limits_{\alpha\in\mathcal{I}_{H}}D_{t}(\xi_{j,\alpha}
\omega_{\alpha}^{(1)})\equiv\chi_j.
\label{rt1}
\ea
$$
It is clear that $\psi_j$ are local functions and that
$D_{t}(\psi_{j})=D(\chi_{j})$. If we show that $\chi_j$ are local as well,
then $D^{-1}(\psi_{j})$ obviously are linear combinations of
$\omega_{\alpha}^{(1)}$ (modulo local functions), so $\tilde{R}_{j}$
are weakly nonlocal UAC functions, 
and the result follows.\looseness=-1 

As $\chi_j$ may depend on the nonlocal variables of the first
level $\omega_{\alpha}^{(1)}$ at most, we only have to check
that $\p\chi_{j}/\p\omega_{\alpha}^{(1)}=0$, i.e.,
$\p\zeta_{j}/\p\omega_{\alpha}^{(1)}=D_{t}(\xi_{j,\alpha})$. 
%
%
Obviously, the only nonlocal terms in $\zeta_{j}$ are\looseness=-1
$$\sum\limits_{\alpha\in\mathcal{I}_{H}}\bigg(\sum\limits_{i=1}^{n}
\sum\limits_{m=0}^{i-1}(-D)^{m}(
(\p\F/\p\bu_{i})^{T}\boldsymbol{\gamma}_{j})D^{i-m-1}(\bH_{\alpha})\bigg)\,
\omega_{\alpha}^{(1)}.$$
Hence, in order to prove our result it remains to 
show that 
$$
\zeta_{j,\alpha}\equiv\sum\limits_{i=1}^{n}
\sum\limits_{m=0}^{i-1}(-D)^{m}(
(\p\F/\p\bu_{i})^{T}\boldsymbol{\gamma}_{j})D^{i-m-1}(\bH_{\alpha})=
D_{t}(\xi_{j,\alpha}).
$$

Comparing this equality with (\ref{lagr3}) and bearing in mind that
$\bH_{\alpha}$ are local generalized symmetries of (\ref{eveq}),
we see that $$
D(\zeta_{j,\alpha})=D_{t}(\boldsymbol{\gamma}_{j}\bH_{\alpha})=
D_{t}(D(\xi_{j,\alpha}))=D(D_{t}(\xi_{j,\alpha})). $$ 
Using Proposition~\ref{prokerd}
from Appendix A, 
we find 
$$
\zeta_{j,\alpha}=c_{j,\alpha}(t)+
D_{t}(\xi_{j,\alpha}), $$ where $c_{j,\alpha}(t)$ is arbitrary
function of $t$.

But it is clear that the function $\xi_{j,\alpha}$, 
determined from 
the relation $D(\xi_{j,\alpha})=
\boldsymbol{\gamma}_{j}\bH_{\alpha}$, is 
defined only up to the
addition of arbitrary element of $\ker D$, i.e., an arbitrary function
$b_{j,\alpha}(t)$ of $t$. 
Hence, replacing $\xi_{j,\alpha}$
by $\xi_{j,\alpha}-\int\limits_{t_{0}}^{t}c_{j,\alpha}(\tau)d\tau$,
we can assume without loss of generality that $c_{j,\alpha}(t)=0$,
and thus $\zeta_{j,\alpha}=D_{t}(\xi_{j,\alpha})$, as required.
\looseness=-1

Thus, we have proved the following result, 
generalizing Proposition~4.1 from \cite{khor}:\looseness=-1
\begin{prop}\label{locrecop}
Let (\ref{eveq}) have a recursion operator $\mathfrak{R}$
(\ref{recop}), where $a_{i}$,$\G_{j}$,$\boldsymbol{\gamma}_{j}$
are local. 
%
Then for any 
$\bH\in\mathrm{WNLS}_{F}(\mathcal{U})$
the quantity $\tilde{\mathfrak{R}}(\bH)$ is well defined and
$\tilde{\mathfrak{R}}(\bH)\in\mathrm{WNLS}_{F}(\mathcal{U})$, 
provided
$\tilde{\mathfrak{R}}(\p\bH/\p\omega_{\alpha}^{(1)})$ are well-defined {\em
local} generalized symmetries of~(\ref{eveq}) for all
$\alpha\in\mathcal{I}_1$. 
\end{prop}

If $\bH =\tilde{\mathfrak{R}}(\G)$ for some local generalized symmetry $\G$,
then $\bH_{\alpha}$ in (\ref{weaknl}) are linear combinations of the
symmetries
$\G_{j}$ that enter into $\mathfrak{R}$.
We can
easily see 
that 
the coefficients
$\tilde\bH_{\alpha}=\p\tilde{\bH}/\p\omega_{\alpha}^{(1)}$ at
$\omega_{\alpha}^{(1)}$ in the representation (\ref{weaknl}) for
$\tilde\bH=\tilde{\mathfrak{R}}^{2}(\G)$ are linear combinations of
$\tilde{\mathfrak{R}}(\bH_{\alpha})$ and of
$\G_{j}$, and hence are in fact linear combinations of $\G_{j}$
and $\tilde{\mathfrak{R}}(\G_{j})$ only.
Thus, 
$\tilde\bH\in\mathrm{WNLS}_{F}(\mathcal{U})$, provided
$\tilde{\mathfrak{R}}(\G_{j})$ are well-defined local generalized symmetries
of (\ref{eveq}). 
This is {\em equivalent} (cf.\ above) to the requirement that 
$\boldsymbol{\gamma}_{j}\tilde{\mathfrak{R}}(\G_{i}) =D(\xi_{i,j})$, where
$\xi_{i,j}$ are local functions, for all $i,j=1,\dots,p$.
\looseness=-1

Iterating this reasoning, we conclude that 
$\tilde{\mathfrak{R}}(\mathbf{Q})$ can be represented in the form
(\ref{recop}) for any nonlocal UAC symmetry $\mathbf{Q}$ obtained by the
repeated application of the recursion operator
$\tilde{\mathfrak{R}}$  to local generalized symmetries, 
provided
$\boldsymbol{\gamma}_{j}\tilde{\mathfrak{R}}^{d}(\G_{i})
=D(\xi_{i,j,d})$, where $\xi_{i,j,d}$ are local functions, for
all $i,j=1,\dots,p$ and all $d=0,1,2,3,\dots$. 
These conditions
are equivalent 
(see above) 
to the requirement that
$\tilde{\mathfrak{R}}^{d}(\G_{i})$ are well-defined {\em local} generalized
symmetries of (\ref{eveq})  for all
$i=1,\dots,p$ and $d=1,2,3,\dots$.\looseness=-1

In particular, if these conditions are satisfied,
then for any (time-dependent) local generalized symmetry $\G$ of
(\ref{eveq}) we have
$\tilde{\mathfrak{R}}^{j}(\G)\in\mathrm{WNLS}_{F}(\mathcal{U})$ for all
$j=1,2,\dots$. Hence, if the
hereditary algebra 
(see e.g.\ \cite{blaszak} for its definition) of time-dependent symmetries
for (\ref{eveq}) is generated by the repeated application of the extension
$\tilde{\mathfrak{R}}$ of a recursion operator $\mathfrak{R}$ (\ref{recop})
to some local generalized symmetries of (\ref{eveq}), and
$a_i$, $\G_{j}$ and $\boldsymbol{\gamma}_{j}$ are local and satisfy the
above conditions, then
all elements of 
this algebra  
belong to
$\mathrm{WNLS}_{F}(\mathcal{U})$.\looseness=-1 

\section{Generalization to non-evolution systems}\label{sec4}
The above results can be applied to any
$(1+1)$-dimensional systems of PDEs transformable into the
evolutionary form (\ref{eveq}) by the appropriate change of
variables.
%
This set includes, in particular, all systems
transformable into Cauchy--Kovalevskaya form\looseness=-1
\be\label{kk0}
{\ds\frac{\p^{r_{I}}u^{I}}{\p
t^{r_{I}}}}=\Phi_{I}\big(x,t,u^{1},\dots,u^{q},\dots,\p^{\alpha+\beta}u^{J}/\p
t^{\alpha}\p x^{\beta},\dots\big),\ \
I=1,\dots,q,
\ee
where $\Phi_{I}$ may depend only on $x,t,u^{1},\dots,u^{q}$ 
and 
$$
\left\{ \p^{\alpha+\beta}u_{J}/\p
t^{\alpha}\p x^{\beta}|\alpha\leq
r_{J}-1,\beta\leq k \right\},\ \ J=1,\dots,q.
$$
The system (\ref{kk0}) 
can be further transformed into an
evolution system of the form (\ref{eveq}) by introducing new
dependent variables $v_{\alpha}^{I}=\p^{\alpha} u^{I}/\p t^{\alpha}$
for $\alpha=1,\dots,r_{I}-1$. Indeed, combining the variables
$u^{1},\dots,u^{q}$ and $v_{\alpha}^{I}$ into a single vector $\mathbf{v}$,
we see that (\ref{kk0}) together with the equations
$$
\p^{\alpha} u^{I}/\p
t^{\alpha}=v_{\alpha}^{I},\,\, \alpha=1,\dots,r_{I}-1, I=1,\dots,q,
$$
forms the evolution system of exactly the same form as
(\ref{eveq}):
\be\label{kk1}
\p\mathbf{v}/\p t=\mathbf{K}(x,t,\mathbf{v},\p\mathbf{v}/\p x,
\p^{2}\mathbf{v}/\p x^{2},\dots,\p^{k}\mathbf{v}/\p x^{k}).
\ee

The class of 
systems of PDEs transformable into
the form (\ref{kk0}) and hence into the evolutionary form
(\ref{kk1}) is very large. In particular, it includes \cite[Ch.\ 2]{o} all
analytic locally solvable $(1+1)$-dimensional systems of PDEs possessing at
least one noncharacteristic direction.
The majority of known examples of non-evolutionary integrable
$(1+1)$-di\-men\-si\-on\-al systems are indeed transformable into the form
(\ref{kk0}) by the (appropriate modification of) 
above change of variables.

Hence, Khor'kova's conjecture stating that
the (extended) recursion operators are well defined on nonlocal UAC
symmetries holds true not only for the evolution systems~(\ref{eveq}) with
the recursion operators (\ref{recop}), but also for any systems
transformable into Cauchy--Ko\-va\-lev\-ska\-ya form (\ref{kk0}) and
then into (\ref{kk1}), provided the recursion operator for
transformed system (\ref{kk1}) has the form (\ref{recop}).
Indeed, making the inverse change of variables we can readily
see that the (extended) recursion operator of original system is also well
defined on its nonlocal UAC symmetries.\looseness=-1

For instance, it is well known that
the sine-Gor\-don equation $u_{\xi\eta}=\sin
u$ can be transformed into $u_{tt}-u_{xx}=\sin u$ by setting
$x=\xi-\eta$, $t=\xi+\eta$. 
Then, introducing a new
dependent variable $v=u_{t}$, we obtain the evolution system 
(see e.g.\ \cite{wang} and references therein),
equivalent to the SG equation:
\be\label{sgee}
u_{t}=v,\, v_{t}=u_{xx}+\sin u.
\ee

The recursion operator $\mathfrak{R}$ for (\ref{sgee}) 
is (see e.g.\ \cite{wang}) of the form~(\ref{recop}),
so by Proposition~\ref{rect1}
the action of $\tilde\mathfrak{R}$ on nonlocal
UAC symmetries  of (\ref{sgee}) is well defined and leaves 
the space of these
symmetries invariant. Returning to the original variables, we conclude that
the same is true for the recursion operator (rewritten as a recursion
operator in the sense of Guthrie \cite{gut}) of SG equation, so 
we recover the result of Khor'kova \cite{khor}, 
initially obtained by straightforward computation.
\looseness=-1

\section{Applications}\label{sec5}

Consider, for instance, the well known integrable Harry Dym
equation $u_{t}=u^{3}u_{3}$. Its
recursion operator (see e.g.\ \cite{wang}) 
$\mathfrak{R}=u^2 D^{2}-u u_{1} D+u u_{2}+u^{3}u_{3}D^{-1}\circ
u^{-2}$ indeed has the form (\ref{recop}), and 
$u^{-2}$ is a cosymmetry and $u^{3}u_{3}$ is
an (obvious) symmetry for this equation.
Using~(\ref{lagr3}), we find 
that $D_{t}(u^{-2}G)=D(D^{2}(u G)-3 u_{1}D(G))$
for any nonlocal UAC symmetry $G$ of HD equation, and hence
$D^{-1}(u^{-2}G)$ is a nonlocal UAC function.
Thus, by Proposition~\ref{rect1} 
the extension $\tilde{\mathfrak{R}}$ of the above
$\mathfrak{R}$ is well defined on the space of nonlocal UAC symmetries of HD
equation and leaves this space invariant.\looseness=-1


It is possible to give a lot of other examples where 
Proposition~\ref{rect1} 
ensures that the 
(extended) recursion operators are well defined on the space of
nonlocal UAC symmetries and leave it invariant. This fact often allows to
draw a number of useful conclusions. For instance, 
provided the hereditary algebra \cite{blaszak,fu} of time-dependent
symmetries is generated by the repeated application of
the extension $\tilde{\mathfrak{R}}$ of a recursion operator $\mathfrak{R}$
of the form (\ref{recop}) to some {\em local} generalized symmetries, all
elements of this algebra are nothing but nonlocal UAC symmetries.
Moreover, if $\tilde{\mathfrak{R}}^{k}(\G_j)$ are well-defined {\em local}
generalized symmetries for all $k\in\mathbb{N}$ and $j=1,\dots,p$, 
then by Proposition~\ref{locrecop}
all elements of this algebra are weakly nonlocal UAC
symmetries, that is, they depend only on the nonlocal variables
$\omega_{\alpha}^{(1)}$, 
i.e., on
the ``integrals'' of
nontrivial local conserved densities for (\ref{eveq}), and 
are linear in $\omega_{\alpha}^{(1)}$.
The hereditary algebra contains time-dependent symmetries of
arbitrarily high order, hence their directional derivatives are formal
symmetries of arbitrarily high order (see e.g.\ \cite{mik} for definition
of formal symmetry) for (\ref{eveq}).
Thus, the evolution systems (\ref{eveq}) possessing the
hereditary algebra 
have time-dependent formal symmetries of arbitrarily high (and hence of
infinite) order, and the coefficients of these formal symmetries are usually
nonlocal (more precisely, weakly nonlocal) UAC functions.\looseness=-2 
\looseness=-1
 
To conclude, let us mention that it would be very
interesting to generalize the results of the present paper to the 
evolution equations 
with constraints,
introduced in \cite{muks}.

\section*{Acknowledgements}
I am pleased to express deep gratitude
to Dr.~M.~Marvan
for the numerous and highly sti\-mu\-lating discussions 
on the subject of this paper. 
\looseness=-1



\section*{Appendix A: On the structure of $\ker D$}
The aim of this appendix is to
prove 
that the
kernel of operator
$D$ in the space $\mathrm{NL}_{F}(\mathcal{U})$
of nonlocal UAC functions
consists solely
of functions of $t$, cf.\ \cite{muks}. 

Let $\mathcal{A}_0$ 
be the algebra of all scalar local functions
under the 
standard mulplication.\looseness=-1 

We shall call an algebra $\mathcal{A}$
of scalar nonlocal UAC functions (under
standard multiplication)
{\em admissible}, if it has the following properties:
\begin{itemize}
\item{for any locally analytic function $h(y_{1},\dots,y_{p})$ 
and any $a_{j}\in\mathcal{A}$ we have\\
$h(a_{1},\dots,a_{p})\in\mathcal{A}$;}
\item{$\mathcal{A}$ is closed under the action of $D$ and $D_{t}$;}
\item{$\mathcal{A}$ is obtained from the algebra $\mathcal{A}_{0}$
by means of a finite sequence of extensions.}
\end{itemize}
The third property means that there exists a finite chain of
admissible algebras $\mathcal{A}_{0},\allowbreak
\mathcal{A}_{1},\allowbreak\mathcal{A}_{2},\allowbreak\dots,
\mathcal{A}_{m}=\mathcal{A}$ such that $\mathcal{A}_{j}$ is 
generated by the elements of
$\mathcal{A}_{j-1}$ and just one new non\-local variable
$\zeta_{j}=D^{-1}(\eta_{j})$, where $\eta_{j}\in\mathcal{A}_{j-1}$ is such
that $\eta_{j}\not\in\Im D|_{\mathcal{A}_{j-1}}$ and
$D_{t}(\eta_{j})\in\Im D|_{\mathcal{A}_{j-1}}$.

Consider a nonlocal UAC function
$f$. It may depend only on a finite number
$m$ of variables $\omega_{\alpha}^{(j)}$, and it is easy
to see that there exists a minimal (i.e., obtained
from $\mathcal{A}_{0}$ by means of the minimal possible number of
extensions) admissible algebra $\mathcal{K}$
of scalar nonlocal UAC functions, which contains $f$.

It is clear that in order to prove that $f\in\ker D$ implies that
$f$ depends on $t$ only it suffices to prove that $\ker
D|_{\mathcal{K}}$ consists solely of functions of $t$.

In order to proceed, we shall need the following
\begin{lem}\label{extadm}
Let $\mathcal{A}$ be an admissible algebra, 
$\ker D|_{\mathcal{A}}$ consist solely of functions of $t$,
and $\tilde\mathcal{A}$ be the extension of $\mathcal{A}$
obtained by adding the nonlocal variable
$\zeta=D^{-1}(\gamma)$, where $\gamma\in\mathcal{A}$ is such that
$\gamma\not\in\Im D|_{\mathcal{A}}$ and $D_{t}(\gamma)\in\Im
D|_{\mathcal{A}}$.

Then $\tilde\mathcal{A}$ is admissible and
$\ker D|_{\tilde\mathcal{A}}$ also consists solely
of functions of $t$.
\end{lem}

{\bf Remark 1.} The conditions 
$\gamma\not\in\Im
D|_{\mathcal{A}}$ and
$D_{t}(\gamma)\in\Im D|_{\mathcal{A}}$ imply that $\gamma$ is a linear
combination of $\rho_{\alpha}^{(j)}$ 
(modulo
the terms from $\Im D|_{\mathcal{A}}$).\looseness=-1
\medskip

{\bf Remark 2.} This lemma is a natural generalization of
Proposition 1.1 from \cite{muks} to the case of time-dependent
nonlocal UAC functions, and its proof relies on the same ideas.
\looseness=-1
\medskip

{\bf Proof of the lemma.} The admissibility of $\tilde\mathcal{A}$
is obvious from the above, so it remains to describe
$\ker D|_{\tilde\mathcal{A}}$. By definition, the elements of
$\mathcal{A}$ may depend only on a finite number of nonlocal
variables $\zeta_{1},\dots,\zeta_{m}$.

Let $\mathcal{B}_{0}\subset\mathcal{A}$ be the algebra of all locally analytic
functions of $x,t,\bu,\allowbreak\bu_1,\dots,
\bu_{p},\allowbreak\zeta_{1},\dots,\zeta_{m}$, where $p$ is the minimal number
such that $\mathcal{B}_{0}$ contains $\gamma$ and $D(\zeta_{i})$ for
$i=1,\dots,m$. It is straightforward to check that such $p$ does
exist.\looseness=-1


Consider the following chain of subalgebras of $\mathcal{A}$:
$$\mathcal{B}_{j+1}=\{ h\in\mathcal{B}_{j} \mid D|_{\mathcal{A}}(h)
=g \gamma, g\in\mathcal{B}_{j}\},\,j=0,1,2,\dots.$$ 

Any locally analytic function of
elements of $\mathcal{B}_{j}$ obviously belongs to $\mathcal{B}_{j}$.

As $\mathcal{B}_{0}$ is generated by $s(p+1)+m+2$
elements
$x,t,u_{I},u_{I,1},\dots,u_{I,p},\zeta_{1},\dots,\zeta_{m}$,
where $I=1,\dots,s$, we conclude that 
$\mathcal{B}_{j+1}=\mathcal{B}_{j}$ for
$j\geq s(p+1)+m+2$ (cf.\ \cite{muks}). 
Indeed, by construction
$\mathcal{B}_{j+1}\subset \mathcal{B}_{j}$, and hence the functional
dimension $d_{j+1}$ of $\mathcal{B}_{j+1}$ does not exceed that of
$\mathcal{B}_{j}$. If these dimensions coincide, then we have
$\mathcal{B}_{j+1}=\mathcal{B}_{j}$, and otherwise $d_{j+1} \leq
d_{j}-1$. Since $d_{0}=s(p+1)+m+2$, it is clear that
$d\equiv d_{s(p+1)+m+2}\leq 1$ provided $\mathcal{B}_{j+1}\neq
\mathcal{B}_{j}$ for $j=0,\dots,s(p+1)+m+1$. On the other hand,
$d\geq 1$, because in any case $\mathcal{B}_{s(p+1)+m+2}$ contains
the algebra of functions of $t$, and the result follows.
Thus, $\mathcal{B}_{s(p+1)+m+2}$ is the algebra of all locally analytic
functions of some its elements $z_{1},\dots,z_{d}$, i.e., it is
generated by $z_{1},\dots, z_{d}$.\looseness=-2

Let $f=f(x,t,\bu,\dots,\bu_{q},
\zeta_{1},\dots,\zeta_{m},\zeta)\in\ker D|_{\tilde\mathcal{A}}$,
$\p f/\p \zeta\neq 0$. Differentiating the equality $D(f)=0$
with respect to $\bu_{j}$, $j>p$, we readily obtain $\p f/\p\bu_{j}=0$ for
$j>p$. Therefore, $f\in\mathcal{B}_{0}$ for any fixed value of $\zeta$.

We have
\be\label{kerdeq}
D(f)=D|_{\mathcal{A}}(f)+\gamma \p f/\p \zeta= 0.
\ee
But (\ref{kerdeq}) implies that $f\in\mathcal{B}_{1}$
for any fixed value of $\zeta$. Then, again by virtue of
(\ref{kerdeq}), we have $f\in\mathcal{B}_{2}$
for any fixed $\zeta$, and so on.

Thus, $f\in\mathcal{B}_{s(p+1)+m+2}$ for any fixed value of
$\zeta$ and $D|_{\mathcal{A}}(f)\neq 0$. Hence, the operator
$D|_{\mathcal{B}_{s(p+1)+m+2}}$ is nonzero and
$D|_{\mathcal{B}_{s(p+1)+m+2}}=\gamma X$, where $X$ is a
nonzero vector field on the space of variables $z_{1},\dots,z_{d}$.

Let $w\in{\mathcal{B}_{s(p+1)+m+2}}$ be a solution of equation
$X(w)=1$. Then $D|_{\mathcal{A}}(w)=\gamma$, what contradicts the
assumption that $\gamma\not\in\Im D|_{\mathcal{A}}$. The contradiction
proves the lemma. $\square$\looseness=-1

The desired result about $\ker D|_{\mathcal{K}}$ 
readily follows, if we successively apply the above lemma for
$\mathcal{A}=\mathcal{A}_{0}$ and $\tilde\mathcal{A}=\mathcal{A}_{1}$, 
then for $\mathcal{A}=\mathcal{A}_{1}$ and
$\tilde\mathcal{A}=\mathcal{A}_{2}$, and so on, until we see that $\ker
D|_{\mathcal{A}_{m}}$,
$\mathcal{A}_{m}=\mathcal{K}$, consists solely of functions of $t$.

Our reasoning applies to 
any nonlocal UAC function $f$,
so we have proved 
\begin{prop}\label{prokerd}
The kernel of the operator $D$ in the space
$\mathrm{NL}_{F}(\mathcal{U})$
of nonlocal UAC functions consists solely
of functions of $t$.
\end{prop}

From this result it is immediate that the intersection $\ker D\cap \ker
D_{t}$ in the space of nonlocal UAC functions consists solely of constants,
and hence 
by Proposition~1.4 from \cite[Ch.\ 6,\S
1]{v} universal Abelian covering over (\ref{eveq}) is locally
irreducible.

Let us stress that 
the above proposition 
is not valid for the functions that depend on the {\em infinite}
number of variables $\bu_{j}$ and $\omega_{\alpha}^{(k)}$ at once.

Indeed, consider the well-known Burgers equation
$u_{t}=u_{2}+u u_{1}$,
whose only nontrivial local conserved density is $u$ (see e.g.\ \cite{v}
for proof). Let $\psi=\psi(x,t,u,u_{1},\dots)$ be an arbitrary
infinitely differentiable local function.
Then it is straightforward to check that the function
$\Psi=\sum\limits_{j=0}^{\infty}
{\ds \frac{{\Upsilon}^{j}(\psi)\omega^{j}}{j!}}$,
where $\omega=D^{-1}(u)$ and $\Upsilon=-(1/u) D$, belongs to $\ker D$. 
It is clear that 
$\Psi$ depends on an infinite number of variables $u_{j}=\p^{j} u/\p x^{j}$,
$j=0,1,2,\dots$, provided $D(\psi)\neq 0$.

\section*{Appendix B: Lie algebra structure of
$\mathrm{NS}_{F}(\mathcal{U})$} 
In this appendix
we prove that the set $\mathrm{NS}_{F}(\mathcal{U})$ is a Lie algebra
with respect to the so-called Lie bracket (see e.g.\ \cite{blaszak}),
defined as 
\be\label{liebr}
[\G,\bH]=\bH'[\G]-\G'[\bH].
\ee
We employed here the notation 
$f'[\bH]=(df(x,t,\bu+\epsilon\bH,\bu_{1}+\epsilon
D(\bH),\dots)/d\epsilon)|_{\epsilon=0}$ for the directional derivative
of any (smooth nonlocal) function $f$ along $\bH$, see e.g.\ \cite{blaszak}.
The bracket (\ref{liebr}) is obviously skew-symmetric. It satisfies the Jacobi
identity by virtue of properties of the directional derivative, see e.g.\
\cite{blaszak} and references therein.

If $f$ is a local function, then 
$f'[{\bH}]=\sum\limits_{i=0}^{\infty}{\ds \frac{\p f}{\p
\bu_{i}}} D^{i}({\bH})$ is a well-defined
nonlocal UAC function for any $s$-component nonlocal UAC vector function
$\bH$, and $\smash{f'=\sum\limits_{i=0}^{\infty}{\ds
\frac{\p f}{\p \bu_{i}}}D^{i}}$ is a differential operator, cf.\ e.g.\
\cite{s}--\cite{sok}.

If $f$ is a nonlocal UAC function, then we have
\be\label{gat2}
f'[{\bH}]=\sum\limits_{k=0}^{\infty}{\ds
\frac{\p f}{\p \bu_{k}}}
D^{k}({\bH})+\sum\limits_{j=1}^{\infty}\sum\limits_{\alpha\in\mathcal{I}_{j}}
{\ds \frac{\p f}{\p\omega_{\alpha}^{(j)}}}{\omega'}_{\alpha}^{(j)} [{\bH}].
\ee
Hence, in order to show that $f'[{\bH}]$ is a nonlocal UAC
function for any $\bH\in\mathrm{NS}_{F}(\mathcal{U})$, we should
define
${\omega'}_{\alpha}^{(j)} [{\bH}]$ and show that ${\omega'}_{\alpha}^{(j)}
[{\bH}]$ are nonlocal UAC functions.\looseness=-1

From now on we assume that the integration constants arising while computing
$D^{-1}$ according to Definition~\ref{defd-1} 
are chosen so that for any constant $c$
we have $\tilde R\equiv D^{-1}(\tilde P)=c
R\equiv c D^{-1}(P)$, where $\tilde P=c P$ (and $\tilde Q=c Q$).
\looseness=-1 

Using the interpretation of ${\omega}_{\alpha}^{(j)}$ as
$D^{-1} (\rho_{\alpha}^{(j-1)})$, given above
(cf.\ \cite{blaszak,fu,muks}), we set
${\omega'}_{\alpha}^{(j)}[{\bH}]=D^{-1}({\rho'}_{\alpha}^{(j-1)}[{\bH}])$.
The quantities ${\rho'}_{\alpha}^{(j-1)}[{\bH}]$ can be computed
inductively. Indeed, $\rho_{\alpha}^{(0)}$ are local functions, so we know
the formula for ${\rho'}_{\alpha}^{(0)}[{\bH}]$ (see above), and hence we
can evaluate ${\omega'}_{\alpha}^{(1)}[{\bH}]$. Next, as
$\rho_{\alpha}^{(1)}$ involve only $\omega_{\alpha}^{(1)}$ and local
variables $x,t,\bu,\bu_1,\dots$, 
we can find ${\rho'}_{\alpha}^{(1)}[{\bH}]$, using
(\ref{gat2}). Then we find ${\rho'}_{\alpha}^{(2)}[{\bH}]$,
and so on.\looseness=-1

According to our definition of $D^{-1}$, 
in order to guarantee 
that $D^{-1}({\rho'}_{\alpha}^{(j)}[{\bH}])$ are
well defined we have to show
that there exist nonlocal UAC functions $\zeta_{\alpha}^{(j)}$ such that
\be\label{noeth}
D_{t}({\rho'}_{\alpha}^{(j)}[{\bH}])=D(\zeta_{\alpha}^{(j)}).
\ee
As $D_{t}(\rho_{\alpha}^{(j)})=D(\sigma_{\alpha}^{(j)})$, we have 
$(D_{t}(\rho_{\alpha}^{(j)}))'[\bH]\allowbreak=
(D(\sigma_{\alpha}^{(j)}))'[\bH]$. Since $\bH\in
\mathrm{NS}_{F}(\mathcal{U})$, we readily obtain from (\ref{symql})
that 
$(D_{t}(\rho_{\alpha}^{(j)}))'[\bH]\allowbreak=
D_{t}({\rho'}_{\alpha}^{(j)}[\bH])$.
It is also easy to see that
$(D(\sigma_{\alpha}^{(j)}))'[\bH]=
D({\sigma'}_{\alpha}^{(j)}[\bH])$. Hence, 
$D_{t}({\rho'}_{\alpha}^{(j)}[\bH])=D({\sigma'}_{\alpha}^{(j)}[\bH])$,
and (\ref{noeth}) holds, if we take ${\sigma'}_{\alpha}^{(j)}[{\bH}]$ for
$\zeta_{\alpha}^{(j)}$. Using this result, we always can
make a `canonical' choice $Q={\sigma'}_{\alpha}^{(j)}[\bH]$ 
while computing $D^{-1}(P)$ for $P={\rho'}_{\alpha}^{(j)}[\bH]$ according to
Definition~\ref{defd-1}, 
and thus get rid of possible pathologies. 

It is clear from the above that if
$\G,\bH\in\mathrm{NS}_{F}(\mathcal{U})$, then $\G'[\bH]$ and $\bH'[\G]$ are
nonlocal UAC vector functions,
and thus
$[\G,\bH]$ is a nonlocal UAC vector function. Using (\ref{symql}) for $\G$ and
$\bH$, we can easily show that
$[\G,\bH]\in\mathrm{NS}_{F}(\mathcal{U})$, so
we have proved the following
\begin{prop}
The set $\mathrm{NS}_{F}(\mathcal{U})$ is a Lie algebra under the
commutator (\ref{liebr}).
\end{prop}



\begin{thebibliography}{99}
\footnotesize

\bibitem{blaszak} M. B\l aszak, {\it Multi-Hamiltonian Theory of
Dynamical Systems}, Springer, Berlin, 1998.

\bibitem{fu}B. Fuchssteiner, Mastersymmetries, higher order time-dependent
symmetries and conserved densities of nonlinear evolution equations,
{
Progr.\ Theor.\ Phys.} 70 (1983) {1508--1522}.\looseness=-1
\bibitem{fu1}B. Fuchssteiner, Integrable nonlinear evolution
equations with time-dependent coefficients, {
J. Math.\ Phys.} 34
(1993) {5140--5158}.
\bibitem{gut}G.A. Guthrie, Recursion operators and non-local symmetries,
{
Proc. Roy. Soc. Lond. A} 446 (1994) {107--114}.
\bibitem{ibrbook} N.H. Ibragimov, {\it Transformation Groups Applied to
Mathematical Physics},  
Reidel, 
Dordrecht--Boston, 
1985.
\bibitem{khor}N.G. Khor'kova, Conservation laws and nonlocal symmetries,
{
Math. Notes} 44 (1988) {No. 1-2, 562--568}.
\bibitem{mbcf}W.X. Ma, P.K. Bullough, P.J. Caudrey and W.I. Fushchych,
Time-dependent symmetries of variable-coefficient evolution equations
and graded Lie algebras, {
J. Phys.\ A} 30 (1997) {5141--5149}.
\bibitem{s}A.V. Mikhailov, A.B. Shabat and V.V. Sokolov,
The symmetry approach to classification of integrable equations,
in: {\em What is Integrability?}, V.E.~Zakharov ed. Springer, New York,
1991.
\bibitem{mik1}A.V. Mikhailov, A.B. Shabat and R.I. Yamilov, The symmetry
approach to classification of nonlinear equations. Complete lists
of integrable systems, 
Russ. Math. Surveys 42 (1987)
No.~4, 1--63.
\bibitem{mik}A.V. Mikhailov and R.I. Yamilov, Towards classification of
$(2+1)$-di\-men\-si\-on\-al integ\-rable equations. Integrability
conditions I, 
J.~Phys.~A: Math. Gen. 31 (1998)
6707--6715.\looseness=-1
\bibitem{muks} F.Kh. Mukminov and V.V. Sokolov, Integrable evolution
equations with constraints, {
Mat.\ Sb.\ (N.S.)} 133(175) (1987)
{No. 3, 392--414}.
\bibitem{o}P.J. Olver, {\it Applications of Lie Groups to
Differential Equations} (Springer, New York, 1986).
\bibitem{sw} J.A. Sanders and J.P. Wang, On recursion operators,
Technical report WS-538, Vrije Universiteit van Amsterdam, 2000, to appear
in {
Physica D}.
\bibitem{sok}V.V. Sokolov, On the symmetries of evolution equations,
Russ. Math. Surveys 43 (1988) No.~5, 165--204.
\bibitem{svin}
S.I. Svinolupov and V.V. Sokolov,
Weak nonlocalities in evolution equations,
{
Math.\ Notes} 48 (1990), No. 5-6, 1234--1239.
\bibitem{v}{\em Symmetries and Conservation Laws for Differential Equations
of Mathematical Physics}, I.S. Krasil'shchik and A.M. Vinogradov eds.,
Translations of Mathematical Monographs 182,
American Mathematical Society, Providence, 
1999.
\bibitem{v0} A.M. Vinogradov, I.S. Krasil'shchik, V.V. Lychagin,
{\it Introduction to Geometry of Nonlinear Differential
Equations}, Nauka, Moscow, 1986. \looseness=-1
\bibitem{wang}J.P. Wang, Symmetries and conservation laws of evolution
equations, Ph.D. Thesis, Vrije Universiteit van Amsterdam, 1998.
\end{thebibliography}
\end{document}